\documentclass[conference,10pt]{IEEEtran}
\IEEEoverridecommandlockouts

\normalsize
\usepackage{bbm}
\usepackage{adjustbox}
\usepackage{enumitem}
\usepackage{cite,algorithm,algorithmic,amsmath,amssymb,amsthm,empheq,mhsetup}
\usepackage{subfigure,amsfonts,balance}
\usepackage{epstopdf}
\usepackage{enumitem}
\usepackage{setspace}
\usepackage[dvipsnames]{xcolor}
\usepackage{amsmath}
\usepackage{array}
\usepackage{amsfonts,amssymb}
\usepackage{bm}
\usepackage{multirow}
\usepackage{tikz}
\usetikzlibrary{decorations.pathreplacing,calligraphy}
\usepackage{pgfplots}
\usepackage[left=0.63in, right=0.63in, top=0.62in, bottom=0.95in,centering]{geometry}

\DeclareMathOperator{\EEE}{\mathbb{E}}

\DeclareMathOperator{\C}{\mathbb{C}}
\DeclareMathOperator{\NNN}{\mathbb{N}}

\DeclareMathOperator{\h}{\mathbf{h}}

\DeclareMathOperator{\FF}{\mathcal{F}}

\DeclareMathOperator{\vv}{\mathbf{v}}
\DeclareMathOperator{\w}{\mathbf{w}}
\DeclareMathOperator{\z}{\mathbf{z}}

\DeclareMathOperator{\RRR}{\mathbb{R}}

\DeclareMathOperator{\CN}{\mathcal{CN}}

\DeclareMathOperator{\rr}{\mathbf{r}}
\DeclareMathOperator{\x}{\mathbf{x}}

\DeclareMathOperator{\y}{\mathbf{y}}

\DeclareMathOperator{\n}{\mathbf{n}}

\DeclareMathOperator{\g}{\mathbf{g}}

\DeclareMathOperator{\THeta}{\boldsymbol{\theta}}

\DeclareMathOperator{\MU}{\boldsymbol{\mu}}

\DeclareMathOperator{\ZETA}{\boldsymbol{\zeta}}

\DeclareMathOperator{\SE}{\mathtt{SE}}
\DeclareMathOperator{\EEEE}{\mathtt{EE}}

\allowdisplaybreaks

\begin{document}
\bstctlcite{IEEEexample:BSTcontrol}
\fontsize{10}{12}\rm

\title{
Joint Optimization of Switching Point and Power Control in Dynamic TDD Cell-Free Massive MIMO

\vspace{-2mm}
\thanks{This work was partially supported by the Wallenberg AI, Autonomous Systems and Software Program (WASP) funded by the Knut and Alice Wallenberg Foundation, partially supported by ELLIIT, and partially by the European Union’s Horizon 2020 research and innovation programme under grant agreement No 101013425 (REINDEER). The simulations were enabled by resources provided by the National Academic Infrastructure for Supercomputing in Sweden (NAISS) at NSC partially funded by the Swedish Research Council through grant agreement no. 2022-06725.}
 }

\author{
\IEEEauthorblockN{
Martin Andersson\IEEEauthorrefmark{1},
Tung T. Vu\IEEEauthorrefmark{1},
Pål Frenger\IEEEauthorrefmark{2},
Erik G. Larsson\IEEEauthorrefmark{1}
}
\IEEEauthorblockA{\small\IEEEauthorrefmark{1}Department of Electrical Engineering (ISY), Link\"{o}ping University, 581 83 Link\"{o}ping, Sweden}
\IEEEauthorblockA{\small\IEEEauthorrefmark{2}Ericsson Research, 583 30 Link\"{o}ping, Sweden}
Email: \{martin.b.andersson, thanh.tung.vu, erik.g.larsson\}@liu.se, pal.frenger@ericsson.com
\vspace{-4mm}
}

\maketitle
\allowdisplaybreaks
\vspace{-0mm}
\begin{spacing}{1}
\begin{abstract}
We consider a cell-free massive multiple-input multiple-output (CFmMIMO) network operating in dynamic time division duplex (DTDD). The switching point between the uplink (UL) and downlink (DL) data transmission phases can be adapted dynamically to the instantaneous quality-of-service (QoS) requirements in order to improve energy efficiency (EE). To this end, we formulate a problem of optimizing the DTDD switching point jointly with the UL and DL power control coefficients, and the large-scale fading decoding (LSFD) weights for EE maximization. Then, we propose an iterative algorithm to solve the  formulated challenging problem using successive convex approximation with an approximate stationary solution. Simulation results show that optimizing switching points remarkably improves EE compared with baseline schemes that adjust switching points heuristically.
\end{abstract}
\end{spacing}

\IEEEpeerreviewmaketitle

\vspace{-0mm}
\section{Introduction}
\vspace{-0mm}
\label{sec:Introd}
To handle the ever-increasing amount of data being sent over wireless networks, more efficient network architectures will be required. One such architecture is cell-free massive multiple-input multiple-output (CFmMIMO), which has emerged as a promising technology to complement the current cellular regime in future wireless networks \cite{Ammar2022CST}. CFmMIMO networks consist of multiple distributed access points (APs) that serve all users (UEs) coherently, providing uniform coverage by avoiding problems with cell-edge UEs. 
Increasing traffic and quality-of-service (QoS) demands together with increasing energy prices raise concerns about the energy consumption in wireless networks \cite{MALMODIN18}. Therefore, it is of importance to design future wireless networks to not only handle the increasing data rates, but also to consider the energy efficiency aspect.

Usually, CFmMIMO networks are assumed to operate in static time division duplex (TDD), where the fraction of time allocated to uplink (UL) and downlink (DL) transmission in each coherence interval is fixed. Static TDD is less likely to adapt to the growing diversity of applications and their wide range of QoS rate requirements. Thus, dynamic TDD (DTDD) has been proposed for CFmMIMO to serve UL and DL UEs more efficiently in each coherence interval \cite{KimCST,kim22VTC}. 
The most basic idea of DTDD is to change the switching point between UL and DL data transmission dynamically. For example, in \cite{kim22VTC}, the authors consider CFmMIMO with DTDD and multiple clusters using a heuristic switching point scheme, which makes the fraction of time allocated to UL/DL proportional to the UL/DL traffic queues. 
A more advanced DTDD approach is network-assisted full-duplex (NAFD) in which UL and DL transmissions are served simultaneously \cite{Razlighi2021TWC,fukue22A,chowdhury2021can}. The authors in \cite{Razlighi2021TWC,fukue22A,chowdhury2021can} consider methods where each AP in a CFmMIMO network can be assigned for either UL transmission, DL transmission, both UL and DL, or be silent, in order to improve the spectral and energy efficiencies. However, NAFD has the drawback of severe cross-link interference (CLI), where the DL UEs will experience interference from the UL UEs, and the UL APs will be interfered by the DL APs. To avoid CLI, which significantly degrades the network performance, we instead focus on optimizing the switching point in this paper. 

\textit{Paper contribution:} 
In this work, we aim to answer the question: \textit{"How much EE gain can be achieved by optimizing the switching point in a DTDD CFmMIMO network?"} To the best of our knowledge, this question has not yet been studied in the literature. First, we formulate a problem that maximizes the EE in a CFmMIMO network by jointly optimizing the DTDD switching point, the UL and DL power control coefficients, and the large-scale fading decoding (LSFD) weights. This problem is constrained by the QoS rate requirements in UL and DL for each UE, and the maximum transmit power at both UE and AP sides. The problem is mixed-integer and highly non-convex with strong coupling among variables. We then propose an iterative algorithm to solve the formulated challenging problem using the successive convex approximation (SCA) technique. We show by numerical results that optimizing the switching point improves the EE of CFmMIMO networks significantly, compared to heuristic DTDD switching point baseline methods including static TDD.

\begin{figure}[t!]
    \centering
    \begin{tikzpicture}[scale=0.8]
        \draw[thick] (0,0) -- (10,0);
        \draw[thick] (0,-1) -- (10,-1);
        \draw[thick] (0,0) -- (0,-1);
        \draw[thick] (10,0) -- (10,-1);
        \draw[thick] (2,0) -- (2,-1);
        \draw[thick] (6,0) -- (6,-1);
        
        \draw [decorate,
        decoration = {calligraphic brace, raise=3pt, amplitude=3pt}, thick] (0.05,0) --  (1.95,0) node[pos=0.5,above=5pt,black]{$\tau_p$};
        \draw [decorate,
        decoration = {calligraphic brace, raise=3pt, amplitude=3pt}, thick] (2.05,0) --  (5.95,0) node[pos=0.5,above=5pt,black]{$\tau_c - \tau_p - \tau_a$};
        \draw [decorate,
        decoration = {calligraphic brace, raise=3pt, amplitude=3pt}, thick] (6.05,0) --  (9.95,0) node[pos=0.5,above=5pt,black]{$\tau_a$};
        \draw [decorate,
        decoration = {calligraphic brace, raise=3pt, amplitude=3pt, mirror}, thick] (0.05,-1) -- (9.95,-1) node[pos=0.5,below=5pt,black]{$\tau_c$};
        \node at (1,-0.5) {UL pilots};
        \node at (4,-0.5) {UL data};
        \node at (8,-0.5) {DL data};
    \end{tikzpicture}
    \vspace{-5mm}
    \caption{Illustration of a DTDD coherence block where the switching point between UL and DL data can be adapted dynamically.}
    \label{adaptiveTDD}
    \vspace{-0mm}
\end{figure}
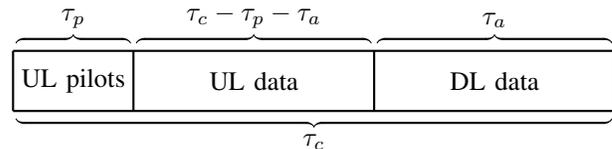

\vspace{-0mm}
\section{System Model}
\label{sec:SystemModel}
We consider a CFmMIMO network with $M$ APs and $K$ UEs. Each AP is equipped with $N$ antennas and the UEs have a single antenna. All APs are connected to the same central processing unit (CPU) via a backhaul network assumed to have a sufficiently high capacity. The APs serve all users simultaneously in the same frequency band using DTDD operation, where each coherence block of $\tau_c$ samples is structured as in Figure \ref{adaptiveTDD}. The first $\tau_p$ samples are used for UL pilot transmission to estimate channels. The DTDD factor $\tau_a \in \NNN$, with $0 \leq \tau_a \leq \tau_c - \tau_p$, corresponding to the allocation of DL transmission, can be changed adaptively to fit the instantaneous UL and DL traffic demands in the network. The APs are assumed to be fully synchronized to allow coherent reception and transmission.

\vspace{-0mm}
\subsubsection{Channel Estimation}
\vspace{-0mm}

Let $\g_{km} \in \C^N$ denote the channel between UE $k$ and AP $m$. We assume uncorrelated Rayleigh-fading channels $\g_{km} = \sqrt{\beta_{km}} \h_{km}$,
where the small-scale fading is represented by $\h_{km} \sim \CN \left( \mathbf{0}, \mathbf{I}_N \right)$ and $\beta_{km} \in \RRR_+$ is the deterministic large-scale fading coefficient. The channel takes a new independent realization in each coherence block. In this paper, the channels are estimated in each coherence block by transmitting UL pilots from the UEs. We assume $K \leq \tau_p$ such that mutually orthogonal pilots can be assigned to all $K$ UEs. At each AP $m$, the channels to all UEs $k$ are estimated using the minimum mean-squared error (MMSE) technique. The channel estimates $\hat{\g}_{km}$ of $\g_{km}$ are based on the received pilot sequences. These estimates have distributions $\hat{\g}_{km} \sim \CN \left( \mathbf{0}, \gamma_{km} \mathbf{I}_N \right)$  \cite{ngo18TGN}, where $\gamma_{km} = \dfrac{\tau_p \rho_p \beta_{km}^2}{\tau_p \rho_p \beta_{km} + 1}, \forall k, m$.
Here, $\rho_p$ denotes the transmission power of the pilots, normalized by the noise power.

\vspace{-0mm}
\subsubsection{Uplink Data Transmission}
\vspace{-0mm}
UE $k$ transmits the signal $x_k = \sqrt{\rho_u \zeta_k} s_k$, where $\rho_u$ is the maximal normalized transmit power from each UE, $\zeta_k$ with $0 \leq \zeta_k \leq 1$ is the UL power control coefficient for UE $k$, and $s_k \in \C$ with $\EEE \{ | s_k |^2 \} = 1$ is the information symbol from UE $k$. AP $m$ receives the signal $\y_m = \sum_{k=1}^K \sqrt{\rho_u\zeta_k} \g_{km} s_k + \n_m$, where $\n_m \sim \CN \left( \mathbf{0}, \mathbf{I}_N \right)$ is the additive receiver noise at the $m$th AP. To detect the transmitted symbols from the individual UEs, we first use maximum ratio combining (MRC) locally at the APs. That is, for each UE $k$ we calculate $\hat{\g}_{km}^H \y_{m}$ at each AP $m$. This is followed by central LSFD to combine the estimates from each AP. For each UE $k$, the locally detected symbols are transferred from all APs to the CPU where they are weighted together to get the final estimate $y_k = \sum_{m=1}^M w_{km} \hat{\g}_{km}^H \y_{m}$, where $w_{km} \in \RRR$, with $0\leq w_{km} \leq 1, \forall k,m$, are the LSFD weights. Finally, $s_k$ is detected from $y_k$. An achievable UL spectral efficiency (SE) for UE $k$ is found by using the use-and-then-forget (UatF) capacity bound \cite{ngo18TGN}:
\begin{align*}
    \SE_k^{\mathtt{ul}} = \dfrac{\tau_c - \tau_p - \tau_a}{\tau_c} \log_2 \left( 1 + \mathtt{SINR}_k^{\mathtt{ul}} \right) \text{ bits/s/Hz},      
\end{align*}
where 
$\mathtt{SINR}_k^{\mathtt{ul}}$ is 
\begin{align}
    \label{SINR_ul}
    \dfrac{N \rho_u \zeta_k \left( \sum_{m=1}^M w_{km} \gamma_{km} \right)^2}{\rho_u \sum_{k'=1}^K \zeta_{k'} \sum_{m=1}^M w_{km}^2 \gamma_{km} \beta_{k'm} \!+ \!\sum_{m=1}^M w_{km}^2 \gamma_{km}}.
\end{align}

\subsubsection{Downlink Data Transmission}

The $m$th AP transmits the signal $\x_m = \sum_{k=1}^K \sqrt{\rho_d} \theta_{km} \vv_{km} q_k$, where $q_k \in \C$ with $\EEE \{ |q_k|^2 \} = 1$ is the information symbol intended for UE $k$, $\rho_d$ is the normalized maximal DL transmit power, and $\vv_{km} \in \C^N$ and $\theta_{km} \geq 0$ is the precoding vector and the power control coefficient for UE $k$ from AP $m$, respectively. The power control coefficients $\theta_{km}$ must fulfil
\begin{align}
    \label{SE1:theta_tot}
    \sum_{k=1}^K \theta_{km}^2 \gamma_{km} \leq \dfrac{1}{N}, \forall m,
\end{align}
to ensure the per-AP power constraint $\EEE \{ \| \x_m \|^2 \} \leq \rho_d$. Here, we only consider maximum ratio (MR) precoding, i.e., $\vv_{km} = \hat{\g}_{km}^{*}, \forall k,m$. Since all APs transmit coherently to all UEs, the $k$th UE receives the signal $r_k = \sum_{m=1}^M \g_{km}^T \x_m + n_k 
= \sum_{m=1}^M \sum_{k'=1}^K \sqrt{\rho_d} \theta_{k'm} \g_{km}^T \hat{\g}_{k'm}^{*} q_{k'} + n_k$, where $n_k \sim \CN \left( 0,1 \right)$ is the normalized receiver noise at UE $k$. Again, we use the UatF capacity bounding technique to derive the following achievable DL SE for UE $k$ \cite{ngo18TGN}:
\begin{align*}
    \mathtt{SE}_k^{\mathtt{dl}} = \dfrac{\tau_a}{\tau_c} \log_2 \left(1 + \mathtt{SINR}_k^{\mathtt{dl}} \right) \text{ bits/s/Hz},
\end{align*}
where
\begin{align}
    \mathtt{SINR}_k^{\mathtt{dl}} = \dfrac{N^2 \rho_d \left( \sum_{m=1}^M \theta_{km} \gamma_{km} \right)^2}{N \rho_d \sum_{k'=1}^K \sum_{m=1}^M \theta_{k'm}^2 \gamma_{k'm} \beta_{km} + 1}.
\end{align}

\subsubsection{Power consumption}
We define the total power consumption $\mathtt{P}^{\mathtt{tot}} $  [W] as \cite{ngo18TGN}
\begin{align}
    \nonumber
    \mathtt{P}^{\mathtt{tot}} = 
   & \dfrac{\tau_c-\tau_p-\tau_a}{\tau_c} \Bigg[ \Xi_u + \sum_{k=1}^K   c_{u,k} \zeta_k  \Bigg] 
   \\ 
   & +  \dfrac{\tau_a}{\tau_c} \Bigg[\Xi_d + \sum_{m=1}^M N \sum_{k=1}^K c_{d,m} \theta_{km}^2 \gamma_{km} \Bigg],
\end{align}
where we have defined $\Xi_u = \sum_{k=1}^K P_{U,k} + \sum_{m=1}^M N P_{\mathtt{cul},m}$ and $\Xi_d = \sum_{m=1}^M N P_{\mathtt{cdl},m} + \sum_{k=1}^K P_{D,k}$. Here, $P_{U,k}$ and $P_{D,k}$ denote the power consumption required to run circuits for transmission and reception at UE $k$, respectively, while $P_{\mathtt{cul},m}$ and $P_{\mathtt{cdl},m}$ denote the power consumption of circuit components at each antenna of AP $m$ for transmission and reception, respectively. Further, let $\sigma_n^2$ be the noise power. We use $c_{u,k} = \dfrac{\rho_u \sigma_n^2}{\chi_{u,k}}$ and $c_{d,m} = \dfrac{\rho_d \sigma_n^2}{\chi_{d,m}}$, where $\chi_{u,k}$ and $\chi_{d,m}$ are the power amplifier efficiencies at UE $k$ and AP $m$. In this work, we choose to combine UE and AP power consumption with equal weights. In the case when the UE power consumption is considered to be more important than the network energy consumption, the power consumption model can easily be extended with scaling weights representing the relative importance given to the UE and network power consumption, respectively, according to some policy.

\section{Problem Formulation and Solution}
\label{sec:problemformulation}
\subsubsection{Problem formulation}
We aim to maximize the EE, i.e., the fraction between the sum SE $\SE^{\mathtt{sum}} = \sum_{k=1}^K \left( \SE_k^{\mathtt{ul}} + \SE_k^{\mathtt{dl}} \right)$, and the power consumption $\mathtt{P}^{\mathtt{tot}}$ by jointly optimizing the dynamic TDD factor $\tau_a$, the UL power control coefficients $\ZETA=\{\zeta_k\}, \forall k$, the DL power control coefficients $\THeta=\{\theta_{km}\}, \forall k,m$, and the LSFD weights $\w=\{w_{km}\}, \forall k,m$. We formulate the following optimization problem:
\begin{subequations}\label{P:SE1}
\begin{align}
    \underset{\tau_a, \ZETA, \THeta, \w}{\max}\,\, &
    \EEEE = \SE^{\mathtt{sum}} / \mathtt{P}^{\mathtt{tot}}
    \\
    \mathrm{s.t.} \,\, 
    \nonumber
    & \eqref{SE1:theta_tot}
    \\
    \label{SE1:tau_a}
    & 0 \leq \tau_a \leq \tau_c - \tau_p
    \\
    \label{SE1:w}
    & 0 \leq w_{km} \leq 1, \forall k,m
    \\
    \label{SE1:zeta}
    &  0 \leq \zeta_k \leq 1, \forall k
    \\
    \label{SE1:eta}
    & \theta_{km} \geq 0, \forall k,m
    \\
    \label{SE1:ul}
    & \SE_k^{\mathtt{ul}} \geq  \overline{\mathtt{SE}}_k^{\mathtt{ul}}, \forall k
    \\
    \label{SE1:dl}
    & \SE_k^{\mathtt{dl}} \geq  \overline{\mathtt{SE}}_k^{\mathtt{dl}}, \forall k.
\end{align}
\end{subequations}
Here, $\overline{\mathtt{SE}}_k^{\mathtt{ul}}$ and $\overline{\mathtt{SE}}_k^{\mathtt{dl}}$ are pre-determined minimal per-UE UL and DL SE QoS rate requirements, respectively. 

\subsubsection{Solution}
\label{sec:problemformulation-solution}
We introduce the new variables $q$, $u$, $\z^{\mathtt{ul}} = \{z^{\mathtt{ul}}_k\}\ , \forall k$, and $\z^{\mathtt{dl}} = \{z^{\mathtt{dl}}_k\}\ , \forall k$, to reformulate problem \eqref{P:SE1} as follows:
\begin{subequations}\label{P:SE2}
    \begin{align}
        \underset{\substack{\tau_a, \ZETA, \THeta, \w, q, u,\\ \z^{\mathtt{ul}}, \z^{\mathtt{dl}}}}{\max}\,\, &
        q
        \\
        \mathrm{s.t.} \,\,
        \nonumber
        & \eqref{SE1:theta_tot}, \eqref{SE1:tau_a} - \eqref{SE1:eta} \\
        \label{SE13:uq}
        & uq \leq \sum_{k=1}^K \left( z^{\mathtt{ul}}_k + z^{\mathtt{dl}}_k \right) \\
        \label{SE2:Ptot}
        & u \geq \mathtt{P}^{\mathtt{tot}} \\
        \label{SE2:ul1}
        & \SE_k^{\mathtt{ul}} \geq         z^{\mathtt{ul}}_k, \forall k \\
        \label{SE2:ul2}
        & z^{\mathtt{ul}}_k \geq \overline{\mathtt{SE}}_k^{\mathtt{ul}}, \forall k \\
        \label{SE2:dl1}
        & \SE_k^{\mathtt{dl}} \geq  z^{\mathtt{dl}}_k, \forall k \\
        \label{SE2:dl2}
        & z^{\mathtt{dl}}_k \geq \overline{\mathtt{SE}}_k^{\mathtt{dl}}, \forall k.
    \end{align}
\end{subequations}
Note that this is a mixed-integer problem, with the integer variable $\tau_a$. We relax it by introducing a continuous variable $t_a$, with $0 \leq t_a \leq 1-\dfrac{\tau_p}{\tau_c}$, which approximates $\dfrac{\tau_a}{\tau_c}$, and solve problem \eqref{P:SE2} with this variable. This relaxation is tight because the value of $\tau_c$ is normally large, e.g., $\tau_c \geq 200$ \cite{ngo18TGN}. 
The relaxed problem is
\begin{subequations}\label{P:SE13}
    \begin{align}
        \underset{\substack{t_a, \ZETA, \THeta, \w, q, u,\\ \z^{\mathtt{ul}}, \z^{\mathtt{dl}}}}{\max}\,\, &
        q
        \\
        \mathrm{s.t.} \,\,
        \nonumber
        & \eqref{SE1:theta_tot}, \eqref{SE1:w} -\eqref{SE1:eta} , \eqref{SE13:uq}, \eqref{SE2:ul2}, \eqref{SE2:dl2}
        \\
        \label{SE3:t_a}
        & 0 \leq t_a \leq 1-\dfrac{\tau_p}{\tau_c}
        \\
        \nonumber
        \label{SE3:Ptot_r}
        & u \geq \left( 1 - \dfrac{\tau_p}{\tau_c} - t_a \right) \Bigg[ \Xi_u + \sum_{k=1}^K   c_{u,k} \zeta_k  \Bigg] 
        \\ 
        &  + t_a \Bigg[\Xi_d + \sum_{m=1}^M  N \sum_{k=1}^K  c_{d,m} \theta_{km}^2 \gamma_{km} \Bigg]
        \\
        \label{SE3:ul}
        & \left( 1 - \dfrac{\tau_p}{\tau_c} - t_a \right) \log_2 \left( 1 + \mathtt{SINR}_k^{\mathtt{ul}} \right) \geq  z^{\mathtt{ul}}_k, \forall k
        \\
        \label{SE3:dl}
        & t_a \log_2 \left( 1 + \mathtt{SINR}_k^{\mathtt{dl}} \right) \geq  z^{\mathtt{dl}}_k, \forall k.
    \end{align}
\end{subequations}

After finding the optimal value $t_a^{\star}$ of $t_a$, we approximate the optimal DTDD factor $\tau_a^{\star}$ by rounding $\tau_c t_a^{\star}$ to its closest integer. However, it is not guaranteed that \eqref{SE2:Ptot}, \eqref{SE2:ul1} and \eqref{SE2:dl1} will be fulfilled after the rounding step. To handle this, we replace \eqref{SE3:Ptot_r}, \eqref{SE3:ul} and \eqref{SE3:dl}, respectively, with
\begin{align}
    \nonumber
    \label{final_u}
    & u \geq \left( 1 - \dfrac{\tau_p}{\tau_c} - t_a + \dfrac{1}{2\tau_c} \right) \Bigg[ \Xi_u + \sum_{k=1}^K  c_{u,k} \zeta_k  \Bigg] 
    \\
    & \qquad + \left( t_a + \dfrac{1}{2\tau_c} \right) \Bigg[\Xi_d + \sum_{m=1}^M N \sum_{k=1}^K  c_{d,m} \theta_{km}^2 \gamma_{km}  \Bigg]
    \\
    \label{final_ul}
    & \left( 1 - \dfrac{\tau_p}{\tau_c} - t_a - \dfrac{1}{2\tau_c} \right) \log_2 \left( 1 + \mathtt{SINR}_k^{\mathtt{ul}} \right) \geq z^{\mathtt{ul}}_k, \forall k
    \\
    \label{final_dl}
    & \left(t_a - \dfrac{1}{2\tau_c} \right) \log_2 \left( 1 + \mathtt{SINR}_k^{\mathtt{dl}} \right) \geq z^{\mathtt{dl}}_k, \forall k.
\end{align}
This will ensure a margin that is equal to the maximal rounding error such that \eqref{SE2:Ptot}, \eqref{SE2:ul1} and \eqref{SE2:dl1} are fulfilled with certainty. To see why, we note that the rounding step implies $\tau_c t_a^{\star} - \dfrac{1}{2} \leq \tau_a^{\star} \leq \tau_c t_a^{\star} + \dfrac{1}{2}$ or equivalently $t_a^{\star} - \dfrac{1}{2\tau_c} \leq \dfrac{\tau_a^{\star}}{\tau_c} \leq t_a^{\star} + \dfrac{1}{2\tau_c}$. Thus, $1 - \dfrac{\tau_p}{\tau_c} - \dfrac{\tau_a^{\star}}{\tau_c} \geq 1 - \dfrac{\tau_p}{\tau_c} - t_a^{\star} - \dfrac{1}{2\tau_c}$ and $\dfrac{\tau_a^{\star}}{\tau_c} \geq t_a^{\star} - \dfrac{1}{2 \tau_c}$, meaning that if the solution $t_a^{\star}$ satisfies \eqref{final_ul} and \eqref{final_dl}, then $\tau_a^{\star}$ will always satisfy \eqref{SE2:ul1} and \eqref{SE2:dl1}. Similarly, one can show that the corresponding property holds for \eqref{final_u} and \eqref{SE2:Ptot}.

To transform problem \eqref{P:SE13} into a more tractable form, we now introduce variables $\MU=\{\mu_k\}, \tilde{\MU}=\{\tilde{\mu}_k\}$, $\forall k$, such that
\begin{align}
	& \mu_{k}^2 \leq \zeta_{k}, \forall k
 	\label{mu}
  \\
  \label{mutilde}
  & \tilde{\mu}_k^2 \geq \zeta_k, \forall k,
\end{align}
as well as $\rr^{\mathtt{ul}} = \{ r_k^{\mathtt{ul}} \}$, $\rr^{\mathtt{dl}} = \{ r_{k}^{\mathtt{dl}} \}, \forall k$, and $\phi$, $\psi$. Using these variables, the problem can be written as follows:
\begin{subequations}\label{P:SE4}
    \begin{align}
        \underset{\substack{t_a, \ZETA, \THeta, \w, q, u, \\ \z^{\mathtt{ul}}, \z^{\mathtt{dl}}, \rr^{\mathtt{ul}} , \rr^{\mathtt{dl}}, \\  \MU, \tilde{\MU}, \phi, \psi}}{\max}\,\, &
        q
        \\
        \mathrm{s.t.} \,\,\,\,\,\,\,\,\,\,\, 
        \nonumber
        & \!\!\!\!\!\!\!\!\eqref{SE1:theta_tot}, \eqref{SE1:w}-\eqref{SE1:eta}, \eqref{SE13:uq}, \eqref{SE2:ul2}, \eqref{SE2:dl2},\eqref{SE3:t_a}, \eqref{mu}, \eqref{mutilde}
        \\
        \label{SE4:u}
        \nonumber
        & u \geq \left( 1 - \dfrac{\tau_p}{\tau_c} - t_a + \dfrac{1}{2\tau_c} \right) (\Xi_u + \phi) 
        \\
        & \qquad + \left( t_a + \dfrac{1}{2\tau_c} \right) (\Xi_d + \psi)
        \\
        \label{SE4:phi}
        & \phi \geq  \sum_{k=1}^K c_{u,k} \zeta_k
        \\
        \label{SE4:psi}
        & \psi \geq  \sum_{m=1}^M N \sum_{k=1}^K c_{d,m} \theta_{km}^2 \gamma_{km}
        \\
        \label{SE4:ul1}
        & \log_2 \left( 1 + \mathtt{SINR}_k^{\mathtt{ul}} \right) \geq  r^{\mathtt{ul}}_k, \forall k 
        \\
        \label{SE4:ul2}
        & \left( 1 - \dfrac{\tau_p}{\tau_c} - t_a - \dfrac{1}{2\tau_c} \right) r^{\mathtt{ul}}_k \geq z^{\mathtt{ul}}_k, \forall k
        \\
        \label{SE4:dl1}
        & \log_2 \left( 1 + \mathtt{SINR}_k^{\mathtt{dl}} \right) \geq  r^{\mathtt{dl}}_k, \forall k 
        \\
        \label{SE4:dl2}
        & \left( t_a - \dfrac{1}{2\tau_c} \right) r^{\mathtt{dl}}_k \geq z^{\mathtt{dl}}_k, \forall k
        \\
        \label{SE4:geq0}
        & \mu_k \geq 0, \tilde\mu_k \geq 0, r^{\mathtt{ul}}_k \geq 0, r^{\mathtt{dl}}_k \geq 0, \forall k, m.
    \end{align}
\end{subequations}
This problem is still difficult to solve due to constraints \eqref{SE13:uq}, \eqref{mutilde}, \eqref{SE4:u} and \eqref{SE4:ul1}-\eqref{SE4:dl2}. To handle this difficulty, we use the SCA approach. Specifically, we find convex approximations of these constraints and solve the convex approximated problem iteratively. 

First, we introduce variables $\mathbf{a} = \{a_{km}\}, \mathbf{b}=\{ b_{kmq} \}$, $\forall k,m, \text{and } q\in \{1,\dots,K\}$, with
\begin{align}
	\label{a}
	& 0 \leq a_{km} \leq \mu_{k} w_{km}, \forall k, m
	\\
	\label{b}
	& \tilde{\mu}_{q} w_{km} \leq b_{kmq}, \forall k,m,q.
\end{align}
Together with \eqref{SE1:w}, \eqref{mu}, and \eqref{mutilde} these constraints imply
\begin{align*}
	& 0 \leq a_{km} \leq \sqrt{\zeta_{k}} w_{km}, \forall k,m
	\\
	& \zeta_q w_{km}^2 \leq b_{kmq}^2, \forall k,m,q.
\end{align*}
By substituting $a_{km}$ into the numerator and $b_{kmq}$ into the denominator of \eqref{SINR_ul}, we get the lower bound $\log_2 (1 + \mathtt{SINR}_{k}^{\mathtt{ul}}) \geq \log_2 \left(1 + \dfrac{A_{k}^2}{B_{k}} \right), \forall k$, where we define $A_{k} = \sqrt{N \rho_u} \sum_{m=1}^M a_{km} \gamma_{km}$, $B_{k} = \rho_u  \sum_{q=1}^K \sum_{m=1}^{M} b_{kmq}^2 \gamma_{km} \beta_{qm} + \sum_{m=1}^M w_{km}^2 \gamma_{km}$.

Also, note that we can write $\log_2 (1 + \mathtt{SINR}_{k}^{\mathtt{dl}}) = \log_2 \left(1 + \dfrac{C_{k}^2}{D_{k}} \right), \forall k$, where $C_{k} = N \sqrt{\rho_d} \sum_{m=1}^M \theta_{km} \gamma_{km}$ and $D_{k} = N \rho_d \sum_{q=1}^K \sum_{m=1}^M  \theta_{qm}^2 \gamma_{qm} \beta_{km}  + 1$. Then, we follow \cite[Eq. (40)]{vu20TWC} to approximate \eqref{SE4:ul1} and \eqref{SE4:dl1} with the following convex constraints:
\begin{align}
	\nonumber
	\label{ul_log_convex_approx}
	&\dfrac{1}{\log 2} \left( \log \left(1+\dfrac{(A_{k}^{(n)})^2}{B_{k}^{(n)}}\right) - \dfrac{(A_{k}^{(n)})^2}{B_{k}^{(n)}} \right.
	\\
	& \left. + 2\dfrac{A_{k}^{(n)}A_{k}}{B_{k}^{(n)}} - \dfrac{(A_{k}^{(n)})^2(A_{k}^2 + B_{k})}{B_{k}^{(n)}((A_{k}^{(n)})^2 +B_{k}^{(n)})} \right) \geq r_{k}^{\mathtt{ul}}
	\\
	\nonumber
	\label{dl_log_convex_approx}
	&\dfrac{1}{\log 2} \left( \log \left(1+\dfrac{(C_{k}^{(n)})^2}{D_{k}^{(n)}}\right) - \dfrac{(C_{k}^{(n)})^2}{D_{k}^{(n)}} \right.
	\\
	& \left. + 2\dfrac{C_{k}^{(n)}C_{k}}{D_{k}^{(n)}} - \dfrac{(C_{k}^{(n)})^2(C_{k}^2 + D_{k})}{D_{k}^{(n)}((C_{k}^{(n)})^2 +D_{k}^{(n)})} \right) \geq r_{k}^{\mathtt{dl}}.
\end{align}

We still need to handle constraints \eqref{SE13:uq}, \eqref{mutilde}, \eqref{SE4:u}, \eqref{SE4:ul2}, \eqref{SE4:dl2}, as well as the new constraints \eqref{a}, \eqref{b}. For this purpose, we follow \cite[Eq. (39)]{vu18TCOM} to get the following convex approximations of \eqref{SE13:uq}, \eqref{mutilde}, \eqref{SE4:u} \eqref{SE4:ul2}, \eqref{SE4:dl2}, \eqref{a}, and \eqref{b}, respectively:
\begin{align}
    \label{uq_convex_approx}
    \nonumber
    & 0.25 \left( (u+q)^2-2(u^{(n)}-q^{(n)})(u-q) + (u^{(n)}\!-\!q^{(n)})^2 \right)  
    \\
    & - \sum_{k=1}^K \left( z^{\mathtt{ul}}_k + z^{\mathtt{dl}}_k \right) \leq 0
    \\
    \label{mutilde_convex_approx}
    &\zeta_k - 2 \tilde{\mu}_k^{(n)} \tilde{\mu}_k + (\tilde{\mu}_k^{(n)})^2 \leq 0
    \\
    \nonumber
    \label{power_convex_approx}
    & -\! u + \left(\! 1 \! - \! \dfrac{\tau_p}{\tau_c} \!+\! \dfrac{1}{2\tau_c} \!\right) (\Xi_u \!+\! \phi) - t_a \Xi_u + \dfrac{1}{2\tau_c} (\Xi_d \!+\! \psi) \!+\! t_a \Xi_d
    \\
    \nonumber
    & +  0.25 \left((t_a-\phi)^2  -2(t_a^{(n)}+\phi^{(n)})(t_a+\phi) + (t_a^{(n)}+\phi^{(n)})^2 \right.
    \\
    &  \!+\! \left. (t_a+\psi)^2 \!- \!2(t_a^{(n)}-\psi^{(n)})(t_a-\psi) + (t_a^{(n)}-\psi^{(n)})^2 \right)  \leq 0
    \\
    \nonumber
    \label{ul_convex_approx}
    & z_{k}^{\mathtt{ul}} - \left( 1 - \dfrac{\tau_p}{\tau_c} - \dfrac{1}{2\tau_c} \right) r_{k}^{\mathtt{ul}} +  0.25 \left((t_a+r_{k}^{\mathtt{ul}})^2 \right. -2(t_a^{(n)}
    \\
    & \left. -(r_{k}^{\mathtt{ul}})^{(n)})(t_a-r_{k}^{\mathtt{ul}}) + (t_a^{(n)}-(r_{k}^{\mathtt{ul}})^{(n)})^2 \right) \leq 0, \forall k
    \\
    \nonumber
    \label{dl_convex_approx}
    & z_{k}^{\mathtt{dl}} \!+\! \dfrac{1}{2\tau_c} r_{k}^{\mathtt{dl}}
    \!+\! 0.25 \left((t_a-r_{k}^{\mathtt{dl}})^2 \!-\! 2(t_a^{(n)}+(r_{k}^{\mathtt{dl}})^{(n)})(t_a+r_{k}^{\mathtt{dl}}) \right.
    \\
    & \left. \qquad + (t_a^{(n)}+(r_{k}^{\mathtt{dl}})^{(n)})^2 \right) \leq 0, \forall k
    \\
    \nonumber
    \label{a_convex_approx}    
    & a_{km} + 0.25 \left((\mu_{k}\!-\!w_{km})^2\!-\!2(\mu_{k}^{(n)}\!+\!w_{km}^{(n)})(\mu_{k}\!+\!w_{km}) \right.
    \\
    & \left. \qquad + (\mu_{k}^{(n)}\!+\!w_{km}^{(n)})^2 \right)
    \leq 0, \forall k, m
    \\
    \nonumber
    \label{b_convex_approx}
    & 0.25 \left((\tilde\mu_{q}+w_{km})^2-2(\tilde\mu_{q}^{(n)}-w_{km}^{(n)})(\tilde\mu_{q}-w_{km}) \right.
    \\
    & \left. \qquad + (\tilde\mu_{q}^{(n)}\!-\!w_{km}^{(n)})^2 \right)
    - b_{kmq} \leq 0, \forall k,m,q.
\end{align}

\begin{figure*}[t!]
  \centering
  \vspace{-0mm}
  \subfigure[Case 1]
  {\includegraphics[width=0.32\textwidth]{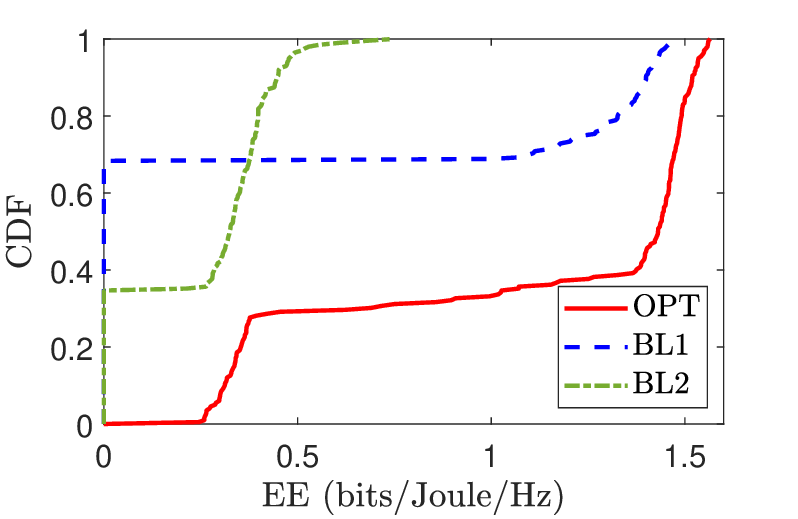}\label{subfig:M30}}
  \vspace{-0mm}
  \subfigure[Case 2]
  {\includegraphics[width=0.32\textwidth]{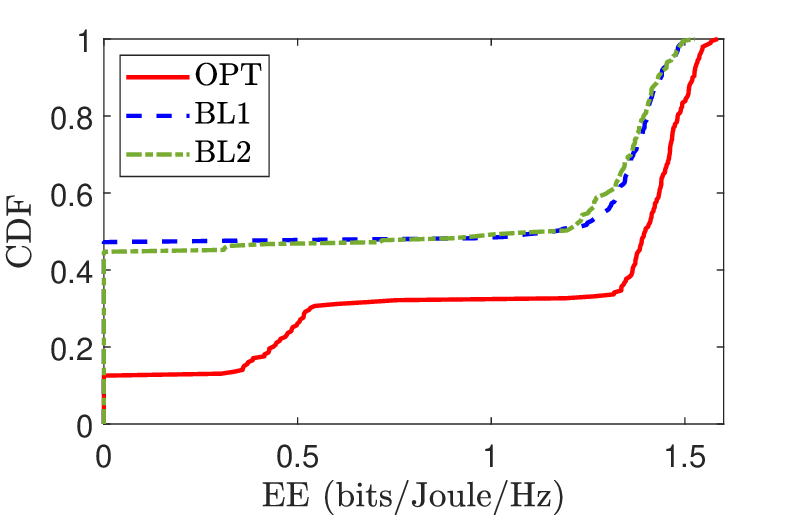}\label{subfig:M60}}
   \vspace{-0mm}
  \subfigure[Case 3]
  {\includegraphics[width=0.32\textwidth]{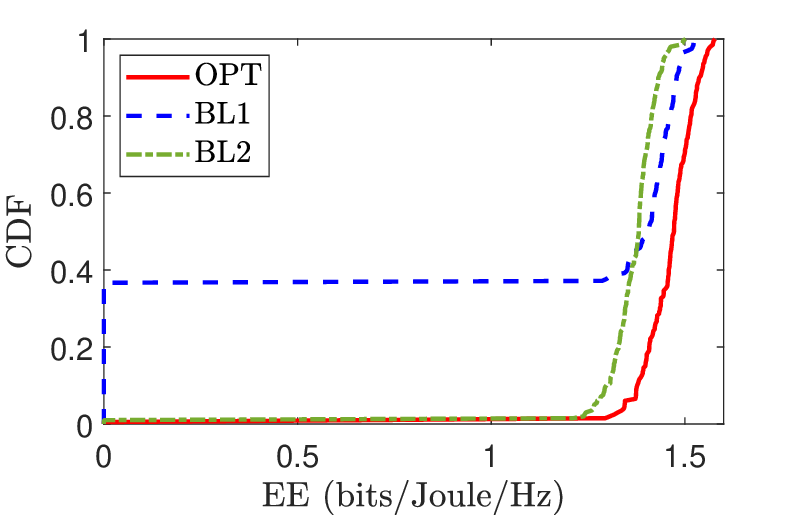}\label{subfig:M60}}
  \vspace{-3mm}
  \caption{Comparisons among the considered schemes.}
  \label{Fig:sim}
  \vspace{-5mm}
\end{figure*}

Now, we can formulate a convex approximation of problem \eqref{P:SE4}:
\begin{align}
	\label{P:SE4_convex}
	\underset{\x}{\max}\,\, &
	q
	\\
	\mathrm{s.t.} \,\, 
	\nonumber
	& \x \in \FF,
\end{align}
where $\x = \{t_a, \ZETA, \THeta, \w, q, u, \z^{\mathtt{ul}}, \z^{\mathtt{dl}}, \rr^{\mathtt{ul}}, \rr^{\mathtt{dl}}, \MU, \tilde{\MU}, \phi, \psi, \mathbf{a}, \mathbf{b} \}$ is the set of all optimization variables, and $\FF = \{
    \eqref{SE1:theta_tot}, \eqref{SE1:w}-
    \eqref{SE1:eta}, \eqref{SE2:ul2}, \eqref{SE2:dl2}, \eqref{SE3:t_a}, \eqref{mu}, \eqref{SE4:phi}, \eqref{SE4:psi}, 
    \eqref{SE4:geq0}, 
    \eqref{ul_log_convex_approx}- 
    \eqref{b_convex_approx}\} $, 
is a convex feasible set. The bounds in \eqref{ul_log_convex_approx}-\eqref{b_convex_approx} are calculated around some fixed point $\x^{(n)}$. To solve problem \eqref{P:SE4}, we use an iterative algorithm that starts at an arbitrary point $\x^{(0)}$ in the convex feasible set $\FF$. Then, we iteratively solve the convex problem \eqref{P:SE4_convex}, where the non-convex constraints are approximated around the solution from the previous iteration, see Algorithm \ref{alg}. The algorithm converges to a Fritz-John solution to problem \eqref{P:SE4}. The proof of this convergence follows \cite{vu18TCOM} and is omitted due to lack of space. Finally, $\tau_a^{\star}$ is calculated from $t_a^{\star}$. This solution is close to a stationary point of the original problem \eqref{P:SE1} due to the tight relaxation of $t_a$ as discussed in Section~\ref{sec:problemformulation-solution}.

\begin{algorithm}[!t]
	\caption{Solution to problem \eqref{P:SE4}}
	\begin{algorithmic}[1]
		\label{alg}
		\STATE \textbf{Initialize}: Set $n\!=\!0$ and draw a random feasible point $\x^{(0)}\!\in\!\FF$.
		\REPEAT
		\STATE Let $n=n+1$.
		\STATE Solve the convex problem \eqref{P:SE4_convex} around $\x^{(n-1)}$ to obtain its optimal solution $\x^*$.
		\STATE Set $\x^{(n)}=\x^*$.
		\UNTIL{convergence}
	\end{algorithmic}
\end{algorithm}

\vspace{-2mm}
\section{Numerical results}
\vspace{-1mm}
The simulations consider a CFmMIMO network where APs and UEs are randomly distributed in a square of $0.5 \times 0.5$ km${}^2$, whose edges are wrapped around to avoid  boundary effects. The distances between adjacent APs are at least $50$ m. The large-scale fading coefficients $\beta_{km}$ are computed using the outdoor 3GPP Urban Microcell model \cite[Eqs. (37), (38)]{emil20TWC}. We set the bandwidth $B=20$ MHz and noise figure $F = 9$ dB. Thus, the noise power is $\sigma_n^2=k_B T_0 B F$, where $k_B=1.381\times 10^{-23}$ Joules/${}^o$K is the Boltzmann constant, while $T_0=290^o$K is the noise temperature. 
We let the maximum UE power be $0.1$ W both for pilot and data transmission, while the maximal power from the APs in the DL is $1$ W. The normalized transmit powers are found by dividing the maximum powers by the noise power $\sigma_n^2$. 
Each coherence block consists of $\tau_c = 200$ samples, whereof $\tau_p = K = 5$ samples are assigned to pilot transmission. 
We choose $M=40, N=1$, $P_{\mathtt{cdl},m} = P_{\mathtt{cul},m} = 0.2$ W, $P_{U,k} = P_{D,k} = 0.1$ W,  $\chi_{u,k} = 0.3$, $\chi_{d,m} = 0.4$ \cite{ngo18TGN}, \cite{bashar19TGCN}.
The following simulation results are averaged over $200$ channel realizations. In each channel realization, if the optimization problem of a scheme is infeasible, we set the EE achieved by that scheme to zero.

To show the effectiveness of our dynamical switching point scheme (\textbf{OPT}), we compare its performance with those of the following heuristic switching point baseline schemes: Static TDD (\textbf{BL1}) where the switching point is set such that the samples for data transmission are divided equally between UL and DL, i.e., $\tau_a = \dfrac{\tau_c - \tau_p}{2}$, and QoS-based DTDD (\textbf{BL2}) where the switching point is set proportionally to the sums of the UL and DL SE QoS requirements, i.e., the number of DL samples $\tau_a$ is the integer closest to $\left( \tau_c-\tau_p \right) \sum_k \overline{\mathtt{SE}}_k^{\mathtt{dl}} \Big/ \sum_k \left( \overline{\mathtt{SE}}_k^{\mathtt{dl}} + \overline{\mathtt{SE}}_k^{\mathtt{ul}} \right)$. Given the chosen switching points, the power control coefficients and LSFD weights of \textbf{BL1} and \textbf{BL2} are optimized using a slightly modified version of Algorithm~\ref{alg}, which is omitted due to the lack of space. 

Fig.~\ref{Fig:sim} compares the EE of all the considered schemes. Three cases of QoS rate requirements of UL in comparison with those of DL transmissions are considered: (Case 1) averagely smaller as in Fig.~\ref{Fig:sim}(a), (Case 2) equal as in Fig.~\ref{Fig:sim}(b), and (Case 3) larger as in Fig.~\ref{Fig:sim}(c).   
Specifically, in Case 1, the QoS rate requirements of each UL and DL UE (in bits/s/Hz) are uniformly chosen as $\overline{\mathtt{SE}}_k^{\mathtt{ul}} \sim U[0.3,0.9]$ and $\overline{\mathtt{SE}}_k^{\mathtt{dl}} \sim U[1.2,1.8]$. In Case 2, $\overline{\mathtt{SE}}_k^{\mathtt{ul}} \sim U[0.5,1.5]$ and $\overline{\mathtt{SE}}_k^{\mathtt{dl}} \sim U[0.5,1.5]$. 
In Case 3, $\overline{\mathtt{SE}}_k^{\mathtt{ul}} \sim U[1.2,1.8]$ and $\overline{\mathtt{SE}}_k^{\mathtt{dl}} \sim U[0.3,0.9]$. New sets of UL and DL QoS rate requirements are generated for each channel realization. 
As seen, \textbf{BL1} and \textbf{BL2} are infeasible for around $50\%$ of all channel realizations in Cases 1 and 2. \textbf{BL1} is infeasible up to $38\%$ of all channel realizations in Case 3. Meanwhile, \textbf{OPT} is infeasible for up to only $12\%$ in Case 2. 
This is reasonable because \textbf{OPT} can adjust the switching point dynamically such that the QoS rate requirements are satisfied. Since the baseline schemes have fixed switching points, they are less likely to adapt to the varying QoS rate requirements of UEs in all channel realizations of the considered cases. 

Fig.~\ref{Fig:sim} also shows the significant advantage of the optimized switching point scheme to improve the EE of DTDD CFmMIMO systems. The median EE of \textbf{OPT} is at least $4.4$ times larger than those of \textbf{BL1} and \textbf{BL2} in Case 1, while being up to $1.3$ times larger than those of \textbf{BL1} and \textbf{BL2} in Cases 2 and 3. More specifically, in Case 1, the median sum SE of \textbf{OPT} is at least $4.6$ times larger than those of \textbf{BL1} and \textbf{BL2}, while the median total power consumption of \textbf{OPT} is at least $3.7$ times smaller than those of \textbf{BL1} and \textbf{BL2}. In Cases 2 and 3,  the median sum SE of \textbf{OPT} is up to $1.8$ times larger than those of \textbf{BL1} and \textbf{BL2}, while the median total power consumption of \textbf{OPT} is similar as those of \textbf{BL1} and \textbf{BL2}. These results are reasonable because optimizing the switching point together with power control and LSFD weights achieves the best trade-off between maximizing the sum rate and minimizing the power consumption in both UL and DL transmission. 
In Cases 2 and 3, since the UL QoS rate requirements are likely much higher than those of Case 1, the UL transmission and power are likely more prioritized. However, since the transmit power of UEs is much smaller than those of APs, the joint optimization of the switching point and power control offers smaller benefits. Thus, the EE gains of \textbf{OPT} over \textbf{BL1} and \textbf{BL2} in Cases 2 and 3 are not as high as that in Case 1.

\vspace{-1mm}
\section{Conclusions}
\vspace{-1mm}
We have investigated the effectiveness of optimizing switching points in terms of
EE in DTDD cell-free massive MIMO. We proposed to maximize EE by jointly optimizing the switching point with the power control coefficients and LSFD weights. Numerical results show that optimizing the switching point is a promising method to improve EE. It offers remarkably higher EE than heuristic switching point baseline schemes. Moreover, it is more likely to give a feasible solution since it is more flexible in adapting to the QoS rate requirements of all UEs. Future research could include, e.g., developing lower complexity algorithms for choosing the switching point.

\vspace{-0mm}

\ifCLASSOPTIONcaptionsoff
  \newpage
\fi

\begin{spacing}{1}
\bibliographystyle{IEEEtran}
\bibliography{IEEEabrv,bib_arxiv}
\end{spacing}

\end{document}